\documentclass[a4paper]{article}
\usepackage{amsmath}
\usepackage{amssymb}
\usepackage{graphicx}
\usepackage{wrapfig}
\usepackage{enumitem}
\graphicspath{{./Image//}}
\usepackage{float}
\usepackage{geometry}
\usepackage{multirow}
\usepackage{tabularx}
\usepackage{rotating}
\usepackage{array}
\usepackage{longtable}
\usepackage{caption}
\usepackage{subcaption}
\usepackage{subfig}
\usepackage{quantikz}
\usepackage{tikz}
\usetikzlibrary{circuits, circuits.ee, positioning}
\usepackage{neuralnetwork}
\usepackage{booktabs}
\usetikzlibrary{positioning}
\usepackage{glossaries}
\makeglossaries

\newacronym{rtdpa}{RTDPA}{Row-Type Dependent Predictive Analysis}
\newacronym{qml}{QML}{Quantum Machine Learning}
\newacronym{qdl}{QDL}{Quantum Deep Learning}
\newacronym{pca}{PCA}{Principal Component Analysis}
\newacronym{ann}{ANN}{Artificial Neural Networks}
\newacronym{smote}{SMOTE}{Synthetic Minority Over-sampling Technique}
\newacronym{roc}{ROC AUC}{Receiver Operating Characteristic Area Under the Curve}
\newacronym{hyquc}{HyQuC-DeepNN-RTDPA}{Hybrid Quantum Classical Deep Neural Network for Row Type Dependent Predictive Analysis }
\newacronym{f1}{F1 Score}{F1 Score (harmonic mean of precision and recall)}

\newcommand{\citep}[1]{\cite{#1}}

\title{Quantum Powered Credit Risk Assessment: A Novel Approach using hybrid Quantum-Classical Deep Neural Network for Row-Type Dependent Predictive Analysis}

\begin{document}
	
	\newgeometry{
		left=3cm, 
		right=3cm, 
			}
	
	\maketitle
	
\author{\begin{center}
		Minati Rath\textsuperscript{*1} and Hema Date\textsuperscript{2}\\
		\textsuperscript{1}Department of Decision Science, IIM Mumbai, India\\
		\textsuperscript{2}Department of Decision Science, IIM Mumbai, India
\end{center}}

\begin{abstract}
\noindent The integration of \gls{qdl} techniques into the landscape of financial risk analysis presents a promising avenue for innovation. This study introduces a framework for credit risk assessment in the banking sector, combining quantum deep learning techniques with adaptive modeling for Row-Type Dependent Predictive Analysis (RTDPA). By leveraging RTDPA, the proposed approach tailors predictive models to different loan categories, aiming to enhance the accuracy and efficiency of credit risk evaluation. While this work explores the potential of integrating quantum methods with classical deep learning for risk assessment, it focuses on the feasibility and performance of this hybrid framework rather than claiming transformative industry-wide impacts. The findings offer insights into how quantum techniques can complement traditional financial analysis, paving the way for further advancements in predictive modeling for credit risk.
\end{abstract}
\section{Introduction}

\noindent In the dynamic landscape of finance, risk assessment is a cornerstone of decision-making, especially in credit lending. While traditional credit risk assessment methods have been somewhat reliable, they often fail to capture the complexities and rapid changes of modern financial markets. This creates a pressing need for advanced tools that improve the speed, accuracy, and efficiency of credit risk analysis \cite{Shi2022}.\\\\Quantum computing, a transformative technology, has demonstrated immense potential across various industries, including finance. Its unparalleled computational capabilities open new avenues for enhancing credit risk assessment processes \cite{Nielsen2000}. This paper introduces a novel framework that integrates quantum computing with classical deep neural networks, referred to as 'Quantum-Powered Credit Risk Assessment,' to address these challenges."\\\\Conventional credit risk assessment techniques rely on historical data, static models, and predefined risk categories, often proving inadequate in adapting to the volatile nature of financial markets and leading to less precise risk predictions. The core objective of this paper is to introduce a more flexible and adaptive methodology through the integration of quantum computing and classical deep neural networks. This approach combines the strengths of quantum computing’s computational efficiency with the learning capabilities of deep neural networks, providing a robust solution to the problem of traditional risk model limitations.\\\\To further enhance the accuracy of credit risk models, this study proposes the use of Row-Type Dependent Predictive Analysis (RTDPA). RTDPA represents a shift in credit risk assessment by recognizing that different loan types exhibit distinct characteristics and risk profiles. By analyzing loan subcategories individually, predictive models can be tailored to the specific attributes of each loan type. This results in more precise risk assessments and a deeper understanding of potential vulnerabilities within a credit portfolio \cite{rath2023adaptivemodellingapproachrowtype}.\\\\The framework for Quantum-Powered Credit Risk Assessment is presented in this paper, and the key research objectives are as follows:
\begin{enumerate}
	\item To develop a hybrid quantum-classical model that enhances the predictive power of credit risk assessments.
	\item To explore how quantum computing can be effectively integrated with classical deep neural networks to improve the analysis of financial data.
	\item To apply the RTDPA technique in credit risk models, improving their adaptability to diverse loan types and facilitating a more granular risk assessment.
\end{enumerate}

\noindent Additionally, the implementation of RTDPA and its benefits for identifying and managing credit risks are discussed, alongside the integration of quantum computing to enhance the performance of deep neural networks. This research also touches upon the potential challenges associated with quantum models, particularly in optimization processes, and introduces alternative optimization techniques like Beetle Antennae Search (BAS), which may offer further improvements for hybrid quantum-classical models. While BAS is not directly applied in this study, it is recognized as a promising avenue for future research.\\\\This research contributes to the intersection of quantum computing, deep learning, and financial risk assessment. By exploring the synergies between these fields, we aim to better understand the potential benefits and challenges of Quantum-Powered Credit Risk Assessment. Ultimately, this approach seeks to offer a more adaptive, precise, and effective framework for credit risk analysis in an increasingly dynamic financial landscape.

\section{Literature Review}

Credit risk assessment is a critical aspect of financial decision-making, central to the lending practices of financial institutions. Traditional methods, including logistic regression and decision trees, have long served as the foundation for credit risk assessment models. These models depend heavily on historical data and predefined risk categories for classifying borrowers into specific risk groups. While these traditional methods have been somewhat effective, they struggle to adapt to the dynamic and complex nature of modern financial markets. This limitation often results in inaccuracies in risk predictions \cite{Marqués20131384}.\\\\In recent years, there has been a notable shift toward exploring more advanced techniques for credit risk assessment. Machine learning, particularly deep learning algorithms, have gained prominence due to their ability to capture intricate patterns and relationships in data. Researchers have investigated the use of neural networks, random forests, and support vector machines for credit scoring \cite{GOLBAYANI2020101251}. These machine learning approaches have shown promise, offering improved predictive accuracy compared to traditional methods \cite{thakkar2024improved} \cite{schuld2019quantum}. However, these models are still limited by their reliance on large amounts of labeled data and can struggle to generalize when faced with novel or unseen scenarios, particularly in dynamic financial environments.\\\\Moreover, quantum computing has emerged as a transformative technology with the potential to reshape various industries, including finance. Quantum computing's ability to perform complex calculations at unprecedented speeds has raised significant interest in its application to financial problems, including credit risk assessment. Quantum algorithms, such as Shor's and Grover's, show great potential for addressing complex financial optimization challenges \cite{Shor2002} \cite{Grover1996}. These quantum algorithms, while powerful, are still in the early stages of development and often require substantial computational resources, which presents a significant challenge for their practical application in real-world financial systems.\\\\ \gls{qml}, an emerging field, investigates the integration of quantum computing with machine learning techniques \cite{jha2020quantum} \cite{farhi2018classification}. QML has the potential to enhance the training and prediction capabilities of machine learning models, offering improved pattern recognition and predictive accuracy in financial applications \cite{lloyd2013quantum} \cite{Schuld2019}. This intersection of quantum computing and machine learning offers exciting opportunities to develop innovative credit risk assessment models. However, the adoption of QML techniques in financial risk models is still limited by issues such as the scalability of quantum systems and the high cost of quantum hardware, making large-scale implementation challenging.\\\\ QDL is an interdisciplinary field that combines the principles of quantum computing with deep learning techniques to develop more powerful machine learning models. This emerging area explores how quantum computing hardware and algorithms can be leveraged to enhance the training and execution of deep neural networks. QDL has the potential to address certain complex problems more efficiently than classical deep learning methods and classical machine learning methods, especially in areas where quantum computation provides an advantage \cite{9528698}. The ability of QDL to handle large datasets and improve convergence rates in deep learning models has been demonstrated in several domains; however, its integration into financial models, particularly for credit risk assessment, remains an area of ongoing research and experimentation.\\\\ In the pursuit of more accurate and efficient credit risk assessment, the integration of quantum computing and classical deep neural networks has garnered attention. This hybrid model combines the computational advantages of quantum computing with the interpretability and stability of traditional classical models \cite{biamonte2017quantum} \cite{havlicek2019supervised}. This synergy between quantum computing and machine learning holds the potential to revolutionize credit risk assessment by harnessing quantum power while addressing the intricacies of diverse loan types. However, most existing research on hybrid quantum-classical models for credit risk assessment lacks a focus on Row-Type Dependent Predictive Analysis (RTDPA), which recognizes that different loan types exhibit distinct characteristics and risk profiles. Our proposed framework aims to fill this gap by leveraging RTDPA within the quantum-powered model to enable more accurate and granular risk assessments for various loan types.\\\\ In conclusion, the literature reviewed here highlights the evolving landscape of credit risk assessment, from traditional methods to advanced machine learning techniques, quantum computing, and innovative approaches like RTDPA. While existing studies have explored various aspects of quantum computing and machine learning for credit risk assessment, our research stands out by integrating quantum computing with RTDPA to provide a more adaptive, precise, and effective framework for credit risk analysis. By presenting a comprehensive overview of these developments and their potential implications, this review lays the groundwork for our proposed research on Quantum-Powered Credit Risk Assessment, which aims to bridge the gap between quantum computing and financial risk assessment.

\section{Deep Learning(DL)}	

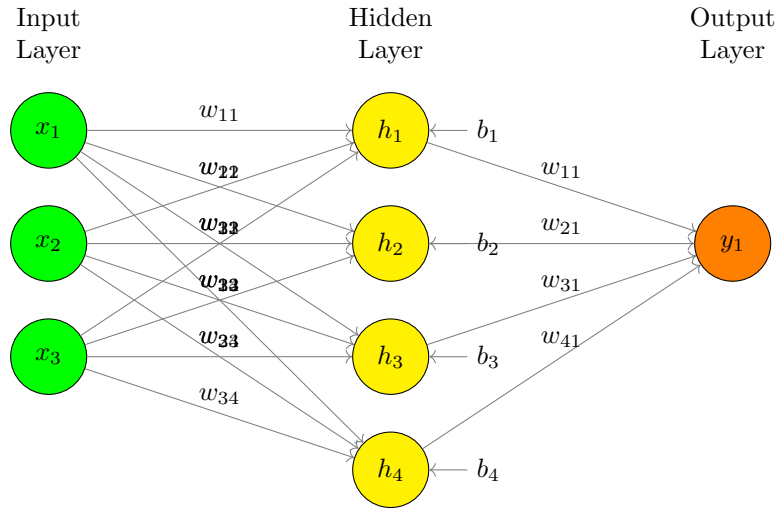
\begin{figure}[H] 
	\centering

	\begin{tikzpicture}[scale=1.5] 
		\foreach \i/\color in {1/green, 2/green, 3/green}
		\node[circle, draw=black, fill=\color, minimum size=1cm] (I\i) at (0, -\i) {$x_{\i}$};
		\node[above=5em of I2, align=center] {Input \\ Layer};
		\foreach \h/\color in {1/yellow, 2/yellow, 3/yellow, 4/yellow}
		\node[circle, draw=black, fill=\color, minimum size=1cm] (H\h) at (3, -\h) {$h_{\h}$};
		\node[above=5em of H2, align=center] {Hidden \\ Layer};
		\node[circle, draw=black, fill=orange, minimum size=1cm] (O1) at (6, -2) {$y_{1}$};
		\node[above=5em of O1, align=center] {Output \\ Layer};
		\foreach \i in {1,...,3}
		\foreach \h in {1,...,4}
		\draw[->, color=gray] (I\i) -- (H\h) node[midway, above, black] {$w_{\i\h}$};
		\foreach \h in {1,...,4}
		\draw[->, color=gray] (H\h) -- (O1) node[midway, above, black] {$w_{\h1}$};
		\foreach \h in {1,...,4} {
			\node[right=0.5cm of H\h, black] (b\h) {$b_{\h}$};
			\draw[->, color=gray] (b\h) -- (H\h);
		}
	\end{tikzpicture}
	\caption{Neural Network Architecture with Weights and Biases}
	\label{fig:NeuralNetwork}
	

	 \begin{minipage}{0.7\textwidth}
		This diagram represents the architecture of a simple feed-forward neural network. The figure consists of three layers:  
		\begin{itemize}
			\item \textbf{Input Layer:} This layer contains 3 input nodes, each representing an input feature (\texttt{$x_{1}$}, \texttt{$x_{2}$}, \texttt{$x_{3}$}).  
			\item \textbf{Hidden Layer:} This layer contains 4 hidden nodes (\texttt{$h_{1}$}, \texttt{$h_{2}$}, \texttt{$h_{3}$}, \texttt{$h_{4}$}). Each hidden node processes input features with weights (\texttt{$w_{\text{input},\text{hidden}}$}) and bias terms (\texttt{$b_{\text{hidden}}$}).  
			\item \textbf{Output Layer:} This layer contains 1 output node (\texttt{$y_{1}$}) that produces the network's prediction. The output is calculated from the weighted sum of the outputs from the hidden layer with weights (\texttt{$w_{\text{hidden},\text{output}}$}).
		\end{itemize}
		The arrows represent the flow of information between the layers, and each connection has an associated weight. Additionally, bias terms are applied to the hidden layer nodes to adjust the output.
	\end{minipage}

\end{figure}

Deep Learning is a sub field of machine learning specialising in the study of neural networks with trainable parameters weight(W) and bias(b).  They are composed of layers of interconnected nodes (neurons). These networks can be deep, meaning they can have multiple hidden layers [Figure \ref{fig:NeuralNetwork}].
 
\[ DL \text{ model} = f_{DL}(x,\theta) = \sigma(Wx+b) , \space \theta=(W,b),\]
with cost function \[ C= \sum_{\text{data}, x, y} |f(x,\theta) - y|^2,\]
Using Gradient Descent $'\nabla'$ of cost 'C' with respect to parameters $'\theta'$ with respect to time 't':
\[\theta^{(t+1)} = \theta^{(t)} - \eta\nabla_\theta C.\] 

\noindent The parameters are updated iteratively in the direction of the gradient until convergence is achieved. After training, the model can be applied to various tasks, such as classification, regression, and image processing.\\\\Deep learning models are particularly powerful because they can autonomously extract hierarchical data representations, identifying pertinent features directly from raw data. This makes them ideal for tasks where traditional feature engineering is either complex or infeasible. However, the effectiveness of these models often depends on the availability of large amounts of labeled data, which can be a significant challenge in certain research domains \\\\ The versatility and adaptability of deep learning algorithms make them applicable across diverse fields. For example, in medical diagnostics, DL models have revolutionized image-based disease detection. Similarly, in the financial domain, deep learning has been leveraged for algorithmic trading and fraud detection. Despite these successes, challenges like interpretability, computational cost, and scalability persist. Understanding the architecture of neural networks is essential to designing efficient and accurate models. The structure of a basic feed-forward neural network is illustrated below, highlighting the flow of data, the role of weights, and the application of biases at various layers. \\\\ The versatility and adaptability of deep learning algorithms make them applicable across diverse fields. For example, in medical diagnostics, DL models have revolutionized image-based disease detection. Similarly, in the financial domain, deep learning has been leveraged for algorithmic trading and fraud detection. Despite these successes, challenges like interpretability, computational cost, and scalability persist. Understanding the architecture of neural networks is essential to designing efficient and accurate models. The structure of a basic feed-forward neural network is illustrated below, highlighting the flow of data, the role of weights, and the application of biases at various layers.\\\\ Beyond traditional applications, deep learning has contributed to advancements in reinforcement learning, enabling breakthroughs in robotics and autonomous systems. These models are also integral to natural language processing, where tasks like sentiment analysis, machine translation, and text generation have seen remarkable progress. The growing availability of pre-trained models and tools has further democratized deep learning, encouraging innovation across multiple domains.
\vspace{1cm}

\vspace{1cm}
\noindent Recent research reveals that deep learning models, with their capacity to automatically learn complex hierarchical patterns and relationships from raw credit data, offer improved predictive accuracy compared to traditional credit scoring methods\cite{Egger20212136} . This enhanced accuracy is attributed to the capability of deep neural networks to capture intricate and non-linear dependencies within diverse sets of features. While these findings underscore the potential of deep learning in credit risk assessment, the literature also highlights challenges related to data availability, model interpretability, and scalability, particularly in the context of vast datasets. The practical applications of deep learning in credit risk prediction extend to risk management and fraud detection, where automated risk assessments can result in more informed lending decisions and improved security against fraudulent activities. Dynamic Deep neural network and recurrent neural network models have shown promising results for portfolio analysis for high frequency trading\cite{0bacd96f632b4c96923264bc498e3ad0}\cite{CAO2023120934}.  Future research should address these challenges and explore specialized architectures tailored to the unique requirements of credit risk modeling, further solidifying deep learning's role in shaping the future of credit risk prediction.
 
\section{Quantum Computing} 
Quantum computing is an advanced computational paradigm that leverages the principles of quantum mechanics to process information. Unlike classical bits, which represent data as either 0 or 1, quantum bits (qubits) can exist in superposition, enabling them to represent multiple states simultaneously. The power of quantum systems arises from the ability to measure complex amplitudes, enabling the manipulation of quantum states in ways that classical systems cannot achieve in a very high dimensional vector space. This property allows quantum computers to perform complex calculations at unprecedented speeds, making them particularly well-suited for algorithms involving large datasets or intricate mathematical operations \cite{rath2023quantumassistedsimulationframeworkdesigning}. In the context of hybrid quantum-classical algorithms, such as those involving Row-Type Dependent Predictive Analysis (RTDPA), quantum computing can enhance feature encoding and model training efficiency \cite{rath2024quantum}. The integration of quantum layers can facilitate improved representation of data patterns, leading to better predictive performance. Furthermore, quantum entanglement allows qubits to be correlated, providing an additional layer of computational advantage by enabling parallel processing of information. As such, quantum computing has the potential to revolutionize machine learning applications, particularly in areas demanding high accuracy and rapid processing capabilities.

\subsection{Quantum Embedding}
Quantum embedding refers to the process of mapping classical data onto a quantum system, enabling quantum algorithms to leverage quantum computing’s advantages. This mapping allows classical data to be represented in quantum states, typically through a process called quantum data encoding. Various techniques exist for quantum embedding, including amplitude encoding, basis encoding, and quantum feature maps, each with its unique advantages and trade-offs. In amplitude encoding, classical data is mapped to the amplitudes of a quantum state, allowing high-dimensional information to be represented efficiently within a smaller number of qubits. The ability to encode classical data in quantum states allows quantum computers to process and manipulate the information in ways that classical systems cannot, making it a valuable technique in the development of hybrid quantum-classical algorithms, such as those used in predictive modeling and machine learning. Quantum embedding, therefore, is an essential component for integrating quantum systems into machine learning workflows, improving both the accuracy and speed of models by exploiting quantum advantages in handling complex and high-dimensional data.

\subsection{Quantum Entanglement}
Quantum entanglement is a fundamental phenomenon in quantum mechanics where qubits become correlated in such a way that the state of one qubit cannot be described independently of the state of the other, even if they are spatially separated. This property allows quantum computers to perform computations that would be impossible for classical computers. When qubits are entangled, they can be used to process information in parallel, dramatically speeding up computations. The entangled qubits share information instantaneously, which can enable faster problem-solving in computationally intensive tasks such as optimization and large-scale data analysis. In the context of quantum machine learning, entanglement plays a critical role by enabling quantum algorithms to efficiently explore and process large search spaces. By leveraging entanglement, quantum algorithms can achieve exponential speedups over classical counterparts in specific tasks, such as solving optimization problems or training complex models. The power of quantum entanglement lies in its ability to create correlations between quantum states that are not possible in classical systems, providing a significant advantage in the processing of large datasets or the solving of otherwise intractable problems.

\section{Quantum Deep Learning}

Quantum Deep Learning combines the properties of quantum computing and deep learning neural networks\cite{biamonte2017quantum}.  Quantum properties like superposition and entanglement, to process information can potentially enhance the capabilities of deep networks for faster convergence of algorithms.  These networks are constructed using quantum gates and qubits instead of classical bits. QDL can take advantage of quantum data encoding, which is the process of encoding classical data in quantum form leading to more efficient processing for certain tasks.  The quantum data can be mapped to a quantum feature space using techniques like the quantum kernel\cite{bravyi2016quantum}. Quantum Neural Networks can be used for tasks like classification, regression, and optimization . \\\\
Moreover, the integration of quantum mechanics into machine learning paradigms opens up new avenues for solving complex problems that are intractable for classical methods. For instance, quantum algorithms can significantly speed up matrix inversion and linear algebra operations\cite{cai2019quantum}, foundational tasks in training deep learning models—enabling models to process larger datasets more efficiently. This enhancement in computational power is particularly beneficial in high-dimensional feature spaces, facilitating improved feature extraction, representation learning, and optimization processes\cite{lourens2023hierarchical}. QDL frameworks can enhance generalization capabilities by utilizing quantum noise and uncertainty, which act as a form of regularization, thereby reducing overfitting in models.\\\\
Recent advances have also demonstrated the potential of quantum neural networks in tasks such as image recognition, where their ability to represent complex patterns offers significant advantages over classical approaches. Furthermore, hybrid architectures that combine classical and quantum methodologies are paving the way for innovative solutions across various domains, including finance—where they assist in credit risk assessment\cite{ORUS2019100028} \cite{Wilkens2023} \cite{Aboussalah2023} \cite{coyle2022quantum}—and healthcare, where they help in predicting patient outcomes and detecting anomalies in medical imaging.
		
\section{Quantum Powered Credit Risk Assessment}

\begin{figure}[H]
	\centering
	\includegraphics[height=10.5cm,width=1.0\textwidth]{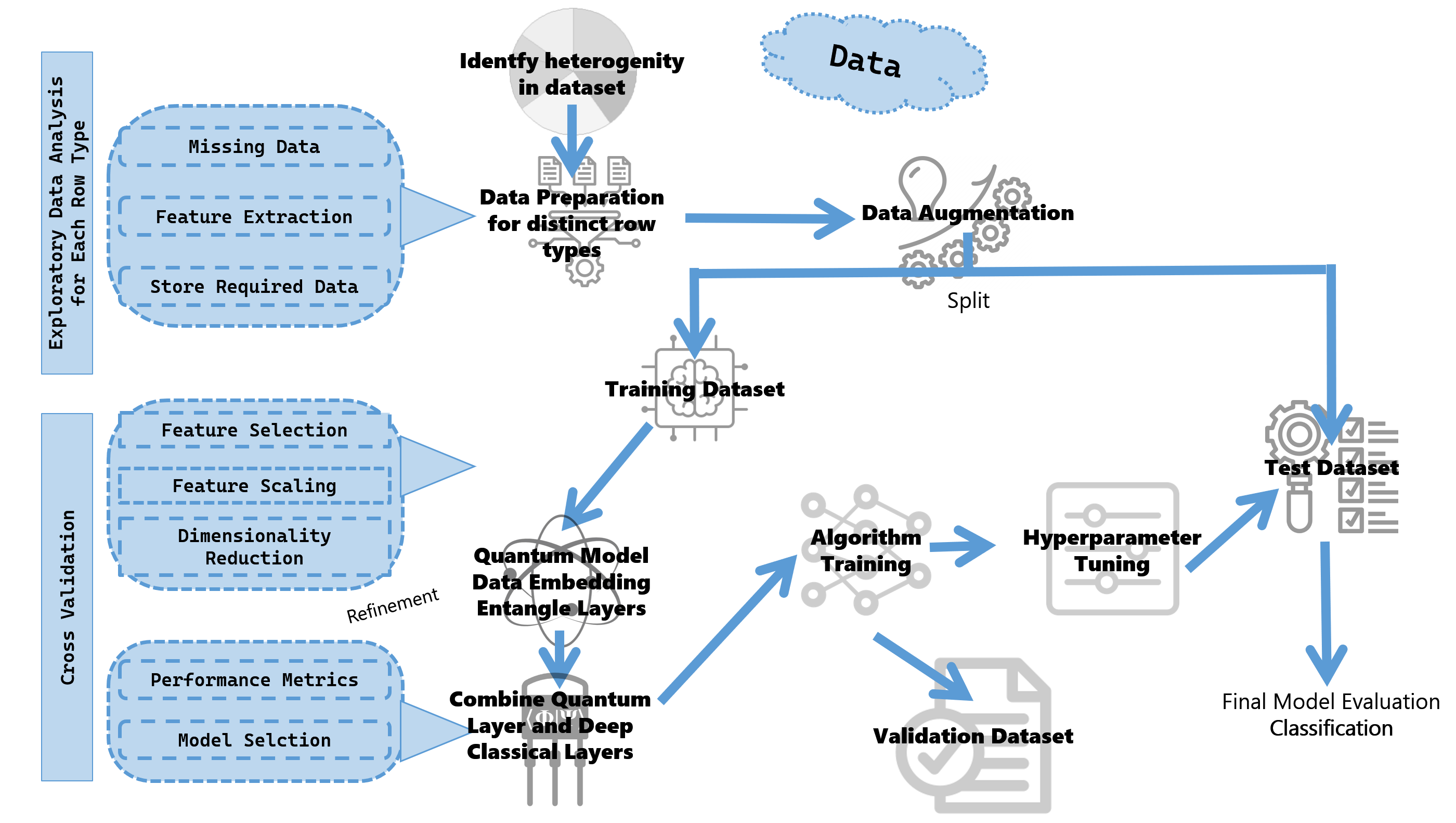}	
	\begin{minipage}{0.85\textwidth}
		\caption{Comprehensive Workflow of the QRTDPA Methodology. This diagram provides a detailed representation of the Quantum Row-Type Dependent Predictive Analysis (QRTDPA) framework, encompassing all critical phases of the experimentation process. Starting with the data flow and identification of heterogeneity in data types, the methodology includes feature engineering, data augmentation, and cross-validation steps. The workflow further incorporates quantum encoding, which transforms classical data into quantum states, followed by the integration of quantum processing layers with classical deep learning layers. The final stages involve algorithm training, hyperparameter optimization, validation on a dedicated dataset, testing on unseen data, and robust model evaluation. This pipeline is designed to synergize quantum and classical approaches for optimal predictive performance.}	
	\label{fig:qrtdpa_methodology}
	\end{minipage}
\end{figure}

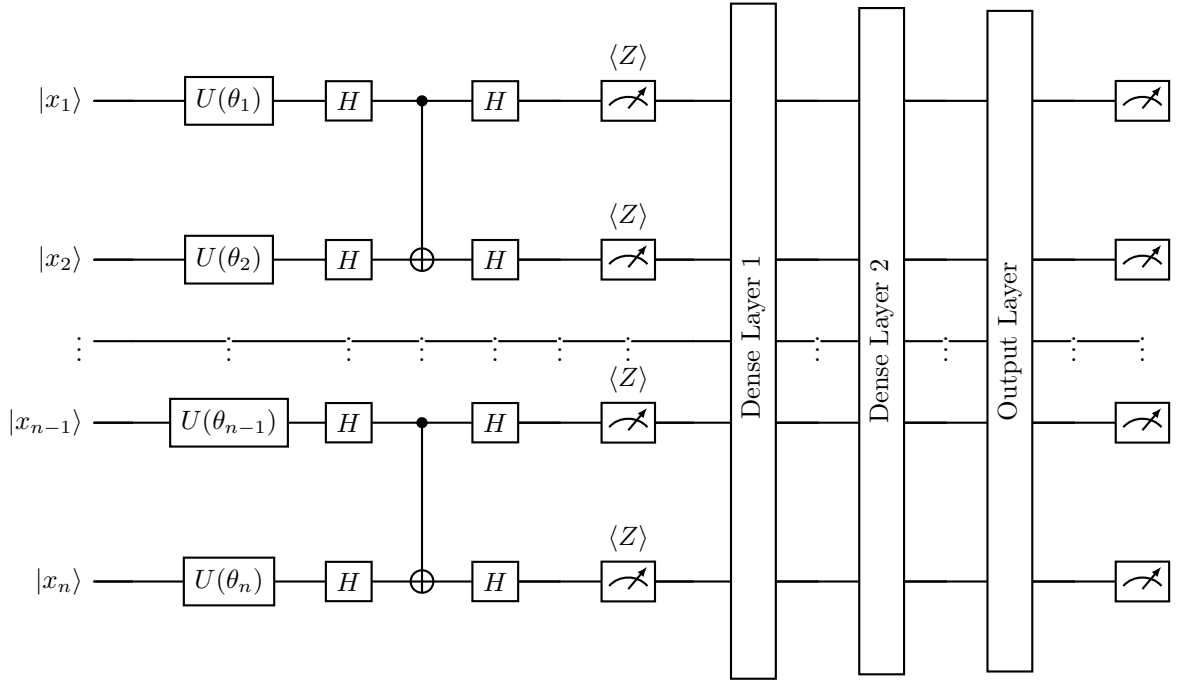
\begin{figure}[H]
	\centering
	\begin{quantikz}
		\lstick{$\ket{x_1}$} & \qw & \gate{U(\theta_1)} & \gate{H} & \ctrl{1} & \gate{H} & \qw & \meter{\langle Z \rangle} & \qw & \gate[wires=5]{\text{\rotatebox{90}{ Dense Layer 1}}} & \qw & \gate[wires=5]{\text{\rotatebox{90}{Dense Layer 2}}} & \qw & \gate[wires=5]{\text{\rotatebox{90}{Output Layer}}} & \qw & \meter{} \\
		\lstick{$\ket{x_2}$} & \qw & \gate{U(\theta_2)} & \gate{H} & \targ{} & \gate{H} & \qw & \meter{\langle Z \rangle} & \qw &  & \qw & & \qw & & \qw & \meter{} \\
		\lstick{$\vdots$} &  & \vdots & \vdots & \vdots & \vdots & \vdots & \vdots &  &  & \vdots &  & \vdots &  & \vdots & \vdots \\
		\lstick{$\ket{x_{n-1}}$} & \qw & \gate{U(\theta_{n-1})} & \gate{H} & \ctrl{1} & \gate{H} & \qw & \meter{\langle Z \rangle} & \qw & & \qw & & \qw & & \qw & \meter{} \\
		\lstick{$\ket{x_n}$} & \qw & \gate{U(\theta_n)} & \gate{H} & \targ{} & \gate{H} & \qw & \meter{\langle Z \rangle} & \qw &  & \qw &  & \qw & & \qw & \meter{}
	\end{quantikz}
	\begin{minipage}{0.7\textwidth}
	\caption{Quantum-Classical Circuit Architecture. The circuit illustrates the hybrid quantum-classical architecture used in this study. The input features ($\ket{x_1}, \ket{x_2}, ..., \ket{x_n}$) are encoded into quantum states via parameterized quantum gates \( U(\theta) \). Quantum processing is followed by measurement operations ($\langle Z \rangle$) to extract classical features, which are fed into dense neural network layers for further processing. The architecture leverages quantum advantages for feature transformation and classical layers for predictive tasks.}
		\label{fig:Hybrid_Quantum_Classical_Circuit}
	\end{minipage}
\end{figure}

This study presents a comprehensive framework for a hybrid quantum-classical deep learning model designed to enhance predictive analysis capabilities, incorporating Row-Type Dependent Predictive Analysis (RTDPA) and Synthetic Minority Over-sampling Technique \gls{smote}. The framework illustrates the integration of quantum computing elements with classical deep learning algorithms, highlighting the synergistic benefits of both approaches [fig \ref{fig:qrtdpa_methodology}]. A detailed algorithm is outlined to guide the implementation of this hybrid model, ensuring efficient data processing and analysis while addressing class imbalance through SMOTE. Additionally, a circuit model is provided, showcasing the quantum and classical components utilized within the framework [fig \ref{fig:Hybrid_Quantum_Classical_Circuit}]. Together, these elements form a robust foundation for advancing hybrid quantum-classical deep learning methodologies in various applications. 

\vspace{0.7cm}

\vspace{0.7cm}
\noindent \textbf{Hybrid Quantum-Classical Deep Neural Network for Row-Type Dependent Predictive Analysis (\gls{hyquc})  Algorithm}\\
		
\textbf{Notation:}
\begin{itemize}
		\item[] $D$: Dataset consisting of $N$ rows and $M$ columns.
		\item[] $R$: Set of distinct row types in $D$ (where $|R|$ is the total number of row types).
		\item[] For each row type $r$ in $R$, $D_r$: Subset of $D$ containing rows of type $r$.
		\item[] $X_r$: Feature matrix for rows of type $r$.
		\item[] $y_r$: Target variable vector for rows of type $r$.
\end{itemize}		
\begin{enumerate}[label=\arabic*.]
		
		\item \textbf{Initialization for Row Type $r$:}
	\begin{align*}
			n_{\text{qubits}}^{(r)} &=\quad n_{\text{wires}}^{(r)} \\
			n_{\text{layers}}^{(r)} &=\quad \text{(number of layers for row type }r\text{)} \\
			n_{\text{classes}}^{(r)} &= \quad \text{(number of classes for row type } r\text{)} \\
			n_{\text{epochs}}^{(r)} &= \quad \text{(number of training epochs for row type } r\text{)} \\
			X_{\text{train}}^{(r)} &= \quad \text{(training input data for row type } r\text{)} \\
			y_{\text{train}}^{(r)} &= \quad \text{(training target labels for row type } r\text{)} \\
			X_{\text{test}}^{(r)} &= \quad \text{(testing input data for row type } r\text{)} \\
			y_{\text{test}}^{(r)} &= \quad \text{(testing target labels for row type } r\text{)} 
	\end{align*}
			
\item \textbf{Data Pre-processing for Row Type $r$:}
	\begin{enumerate}[label=\alph*.]
	\item[] Perform row-type-specific data preprocessing steps, including handling missing values,
	outlier detection, feature engineering, and data transformation specific to row type $r$.
	\item[] Identify the minority class for row type $r$.
	\item[] Apply quantum-inspired or classical data augmentation for class imbalance within row type $r$.
	\item[] Obtain the preprocessed feature matrix $X_r$ and target variable vector $y_r$.
	\end{enumerate}		
\item \textbf{Quantum Model:}
Define quantum model $Q$ using the Quantum Node (QNode):
\begin{align*}
	&\text{Embed data into quantum states:} \\ 
	&\quad \text{Apply a quantum embedding operation to map input data } X_r \text{ into quantum states.} \\
	&\quad \text{This can be achieved using encoding gates, such as Angle Embedding or Amplitude Embedding.} \\
	&\quad \text{Let } \left| \psi_0 \right\rangle = \text{Embed}(X_r). \\
	&\text{Entangle quantum layers:} \\ 
	&\quad \text{Apply a series of entangling gates to create quantum entanglement between qubits.} \\
	&\quad \text{This can be implemented using StronglyEntanglingLayers or custom entangling gates.} \\
	&\quad \text{Let } \left| \psi_{\text{entangled}} \right\rangle = \text{Entangle}(\left| \psi_0 \right\rangle).
	&\text{Apply measurement operators:} \\ 
	&\quad \text{For each qubit, measure the expectation value of a specific observable (e.g., } Z \text{ operator).} \\
	&\quad \text{Compute } \left[ \langle Z_i \rangle \right]_{i=0}^{n_{\text{qubits}}} \text{ using measurements on } \left| \psi_{\text{entangled}} \right\rangle. \\
	&\text{Combine measurement results:} \\ 
	&\quad \text{The resulting measurement outcomes form the quantum model's predictions for each qubit.} \\
	&\quad \text{Let } Q(X_r, \text{weights}) = \left[ \langle Z_i \rangle \right]_{i=0}^{n_{\text{qubits}}}, \text{where $\text{weights}$ are the trainable parameters of the deep quantum model.}
\end{align*}

\item \textbf{Quantum Layer:}
	Define quantum layers $Q_{\text{layer}}$ as part of the deep architecture:
	\[Q_{\text{layer}}(X_r) = Q(X_r, \text{weights})\]
			
	\item \textbf{Classical Model:}
	Define classical deep layers $C_{\text{model}}$ using a Sequential architecture:
	\[ C_{\text{model}}(X) = \sigma(W_{\text{dense1}} \cdot X_r + b_{\text{dense1}}) \]
	\[\text{softmax}(C_{\text{model}}(X_r)) = \text{softmax}(W_{\text{dense2}} \cdot C_{\text{model}}(X_r) + b_{\text{dense2}})\]
				
	\item \textbf{Hybrid Quantum Classical Deep Neural Network Model:}
	Combine the quantum layer and deep classical model to create the  hybrid model:
	\[ H_{\text{model}}^{(r)}(X_r) = \text{softmax}(W_{\text{dense3}} \cdot Q_{\text{layer}}(X_r) + b_{\text{dense3}}) \]
	\item \textbf{Model Training:}
	For each row type $r$ in $R$: 
	\item[] Train the deep hybrid model $H_{\text{model}}^{(r)}$ on the preprocessed data for row type $r$.
	
\item \textbf{Hyperparameter Tuning for Row Type \( r \):}
\hspace*{2em} \begin{enumerate}[label=\arabic*.]
	\item[] Define Hyperparameter Space for Row Type \( r \):
	\begin{itemize}
		\item[] \( n_{\text{layers}} \): Number of quantum layers (e.g., 1, 2, 3)
		\item[] \( n_{\text{qubits}} \): Number of qubits (e.g., 2, 3, 4)
		\item[] \( \text{learning\_rate} \): Learning rates for the classical model (e.g., 0.01, 0.001)
		\item[] \( \text{batch\_size} \): Batch sizes (e.g., 16, 32)
		\item[] \( n_{\text{epochs}} \): Number of training epochs (e.g., 50, 100)
	\end{itemize}	
	\item[] Grid Search Implementation:
	\begin{itemize}
		\item[] Create a grid of hyperparameters by combining the values defined in the hyperparameter space.
		\item[] For each combination of hyperparameters:
		\begin{itemize}
			\item[] Initialize the hybrid model \( H^{(r)}_{\text{model}}(X_r) \) using the current hyperparameter values.
			\item[] Train the model on the preprocessed data \( D_r \) for \( n_{\text{epochs}} \).
			\item[] Evaluate the model using cross-validation on the training data.
		\end{itemize}
	\end{itemize}	
	\item[] Select Best Hyperparameters:
	\begin{itemize}
		\item[] Identify the combination of hyperparameters that resulted in the best performance metric (e.g., accuracy, \gls{f1}-score).
		\item[] Store the best hyperparameters for use in the final model training.
	\end{itemize}

	\item[] Train Final Model:
	\begin{itemize}
		\item[] Initialize the hybrid model \( H^{(r)}_{\text{model}}(X_r) \) using the best hyperparameters.
		\item[] Train the model on the full training dataset \( D_r \) using the best hyperparameter values.
	\end{itemize}	
	\end{enumerate}	
	
\item \textbf{Model Prediction:}
	Perform predictions using the trained deep hybrid model $H_{\text{model}}^{(r)}$ for any new row $x_{\text{new}}$ with an associated row type $r_{\text{new}}$:
	\[\text{predictions} = H_{\text{model}}^{(r_{\text{new}})}.\text{predict}(x_{\text{new}}) \]
	\end{enumerate}

\section{Experimental Results}
In this section, we present description of data, algorithm execution and the results of our analysis.\\

\subsection{Data}
 Our dataset comprises over 25,000 samples sourced from xxx Bank, encompassing 81 attributes detailed in Table\ref{tab:column_description}. Within this dataset, loan types can be distinctly grouped into two categories: Agriculture loans and Personal loans. Each loan entry falls under one of four classifications: standard, substandard, doubtful, or loss, as outlined in Table\ref{tab:Loan Classification}. Notably, for personal loans, there are only three samples categorized as "loss," prompting us to consolidate them into the "doubtful" category. To implement hybrid quantum classical deep neural network for Row-Type Dependent Predictive Analysis (\gls{hyquc}), we have segregated rows associated with each subcategory of loans and pre-processed them individually.
 \vspace{1cm}

\begin{table}[H]
	\centering
	\caption{Loan Classification}
	\label{tab:Loan Classification}
	\begin{tabular}{|l|l|l|}
		\hline
		\textbf{Loan Type} & \textbf{IRAC} & \textbf{Count} \\
		\hline
		\multirow{4}{*}{Agriculture Loan} & Standard & 17496 \\
		\cline{2-3}
		& Sub Standard & 294 \\
		\cline{2-3}
		& Doubtful & 2577 \\
		\cline{2-3}
		& Loss & 210 \\
		\hline
		\multirow{4}{*}{Personal Loan} & Standard & 4398 \\
		\cline{2-3}
		& Sub Standard & 126 \\
		\cline{2-3}
		& Doubtful & 129 \\
		\cline{2-3}
		& Loss & 3 \\
		\hline
	\end{tabular}
	\vspace{1em} 
\begin{flushleft}
	\hspace{8em}
	\textbf{Table \ref{tab:Loan Classification}:} Loan Classification based on IRAC criteria.
	
\begin{itemize}[left=3cm]  
	\item \textbf{Loan Type:} Agriculture or Personal Loan.
	\item \textbf{IRAC:} Loan status categories based on risk:
	\begin{itemize}[left=1cm]  
		\item \textbf{Standard:} Low risk.
		\item \textbf{Sub Standard:} Manageable risk.
		\item \textbf{Doubtful:} High risk of default.
		\item \textbf{Loss:} Expected default.
	\end{itemize}
	\item \textbf{Count:} Number of loans in each category.
\end{itemize}

\end{flushleft}

\end{table}

 \subsection{Data PreProcessing}

Features dryland and wetland are not applicable to personal loan, so we removed them from analysis of personal loan.  We also analysed missing values for each category and dropped features having missing values more then $70\%$ as listed in table\ref{tab:missing-values} below. 

\begin{table}[H]	
	\caption{\% of Missing Values for Personal Loan and Agriculture Loan}
	\label{tab:missing-values}
	\begin{subtable}{0.4\textwidth}
		\centering
		\caption{Personal Loan}
		\begin{tabular}{|lcc|}
			\hline
			Variable   & Total Missing & \% Missing \\
			\hline
			OPINIONDT  & 4436          & 95.3 \\
			DIRFINFLG  & 4656          & 100.0 \\
			SANAUTCD   & 3838          & 82.4 \\
			DOCREVDT   & 4229          & 90.8 \\
			PRISECCD2  & 4406          & 94.6 \\
			RENEWALDT  & 4656          & 100.0 \\
			INSEXPDT   & 4439          & 95.3 \\
			TFRDT      & 4216          & 90.5 \\
			REASONCD   & 4181          & 89.8 \\
			RECALLDT   & 4216          & 90.5 \\
			WOSACD     & 4568          & 98.1 \\
			\hline
		\end{tabular}
	\end{subtable}%
	\begin{subtable}{0.7\textwidth}
		\centering
		\caption{Agriculture Loan}
		\begin{tabular}{|lcc|}
			\hline
			Variable   & Total Missing & \% Missing \\
			\hline
			OPINIONDT  & 19743         & 95.9 \\
			SANAUTCD   & 18067         & 87.8 \\
			DOCREVDT   & 19918         & 96.8 \\
			PRISECCD2  & 19668         & 95.6 \\
			UNIFUNFLG  & 15771         & 76.6 \\
			RENEWALDT  & 19956         & 97.0 \\
			INSEXPDT   & 20502         & 99.6 \\
			TFRDT      & 18090         & 87.9 \\
			REASONCD   & 18081         & 87.9 \\
			RECALLDT   & 18096         & 87.9 \\
			WOSACD     & 19579         & 95.1 \\
			\hline
		\end{tabular}
	\end{subtable}
	
\begin{flushleft}
	\textbf{Table \ref{tab:missing-values}:} Percentage of Missing Values for Personal and Agriculture Loan datasets.
	\begin{itemize}
		\item \textbf{Personal Loan:} Lists variables with missing values and their percentages.
		\item \textbf{Agriculture Loan:} Lists variables with missing values and their percentages.
		\item \textbf{Total Missing:} Total missing values for each variable.
		\item \textbf{\% Missing:} Percentage of missing values, calculated by dividing missing values by total observations.
	\end{itemize}
\end{flushleft}

\end{table}

\subsection{Feature Selection}
\begin{figure}[H]
	\centering
	\begin{subfigure}{0.5\textwidth}
		\centering
		\includegraphics[width=\linewidth]{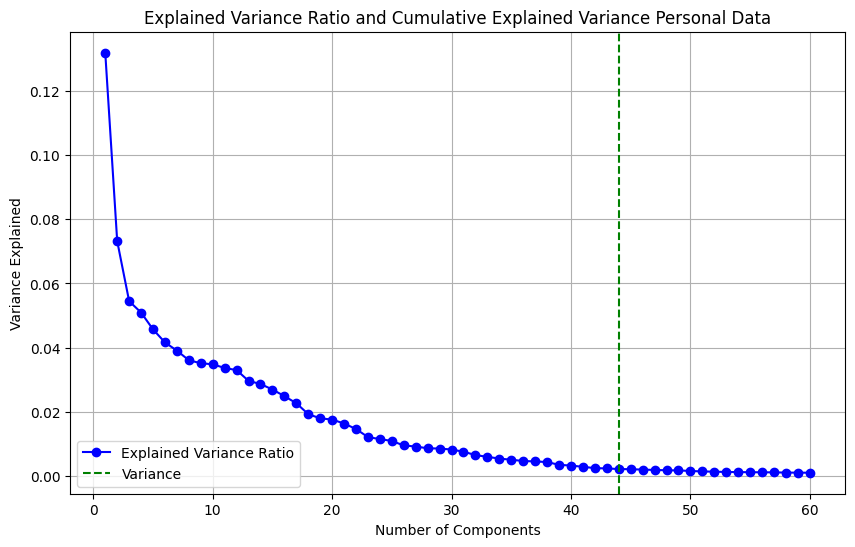} 
		\caption{PCA Personal Data}
		\label{fig:subfig1}
	\end{subfigure}%
	\begin{subfigure}{0.5\textwidth}
		\centering
		\includegraphics[width=\linewidth]{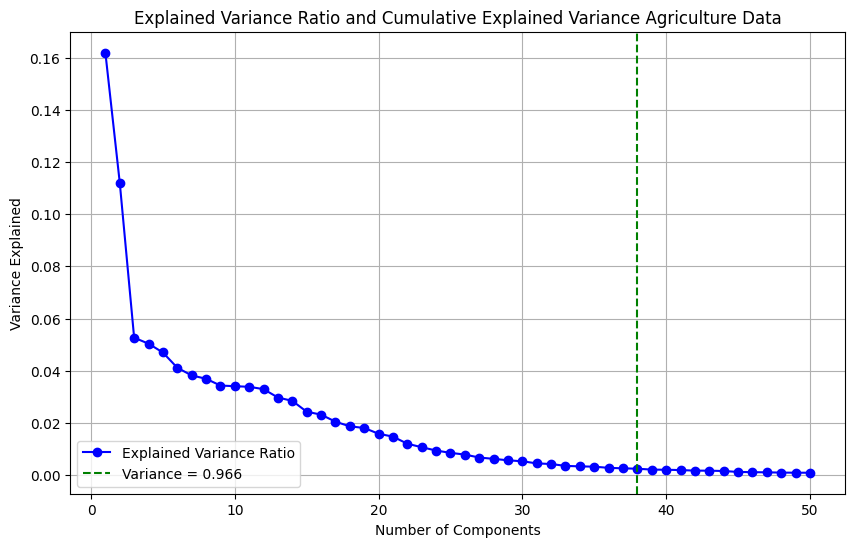} 
		\caption{PCA Agriculture Data}
		\label{fig:subfig2}
	\end{subfigure}
	\caption{PCA}
	\label{fig:mainfig}
	\vspace{1em} 
	 \begin{minipage}{0.8\textwidth} 
		\text{ Principal Component Analysis (PCA) on Personal Loan and Agriculture Loan Data.}
		\begin{itemize}
			\item \textbf{PCA Personal Data (Subfigure \ref{fig:subfig1}):} Visual representation of PCA applied to the personal loan dataset. The plot shows how the data points are distributed across the principal components and the variance explained by each component.
			\item \textbf{PCA Agriculture Data (Subfigure \ref{fig:subfig2}):} Visual representation of PCA applied to the agriculture loan dataset. This plot displays the distribution of the agriculture loan data across the principal components and how much variance each component explains.
			\item \textbf{PCA Objective:} PCA is used here to reduce the dimensionality of the datasets while retaining the maximum variance, helping to identify patterns or relationships between variables.
		\end{itemize}
	\end{minipage}
\end{figure}

After preprocessing the data, we removed features with unique values across categories and handled missing values. To address the challenge of high dimensionality, we employed Principal Component Analysis (\gls{pca})\cite{Chen2016}, a statistical technique often used to examine relationships between variables and reduce dimensionality. PCA projects the data onto a new vector space determined by the eigenvectors of the original dataset, allowing us to evaluate variable importance while reducing redundancy and dimensionality. By generating linear combinations of the original variables, PCA yields a new set of variables that expose clearer underlying patterns within the data. \\\\ To identify the optimal number of principal components capturing most of the data’s variance, we created a scree plot. This plot reveals the “elbow” or “knee” point where the explained variance or eigenvalues begin to level off, indicating the number of components that should be retained to avoid diminishing returns in capturing additional variance. For our model, we selected 43 principal components for personal data \ref{fig:subfig1} and 38 for agricultural data \ref{fig:subfig2}. \\\\ However, due to limitations in quantum simulators, which are currently unable to process data with up to 38 and 43 principal components, we tested our algorithm using only the first 5 principal components. This reduction was necessary to accommodate the computational constraints of quantum simulation while still allowing us to evaluate the algorithm’s effectiveness on a manageable subset of the data. By focusing on the most significant components, we aimed to preserve the essential structure and variance within the data, ensuring meaningful insights despite the reduced dimensionality.

\hspace{1cm}

\subsection{Data Augmentation}

Data augmentation is crucial in machine learning, especially for imbalanced datasets. In the banking sector, where data often reflects significant class imbalance, it helps improve model performance and robustness. This technique is vital for enhancing predictive accuracy in loan assessments\cite{Mahbobi2023}. \\\\ In particular, loan types such as Agriculture and Personal Loans exhibit imbalanced data distributions. For instance, in Agriculture Loans, we observe 17,496 Standard cases versus only 294 Substandard and 210 Loss cases. Similarly, Personal Loans show 4,398 Standard against just 126 Substandard and 3 Loss cases. \\\\ Data augmentation generates synthetic samples for minority classes, allowing models to learn effectively from these underrepresented categories. This leads to better generalization, reduces the risk of over fitting, and ensures compliance with fair lending regulations. By employing data augmentation, banks can improve risk assessment models, thereby enhancing overall loan portfolio performance and making more informed lending decisions\cite{li2021}.

\section{Result Analysis}
\subsection{Personal Loan}

\begin{figure}[H]
	\centering
	\includegraphics[width=1.0\textwidth]{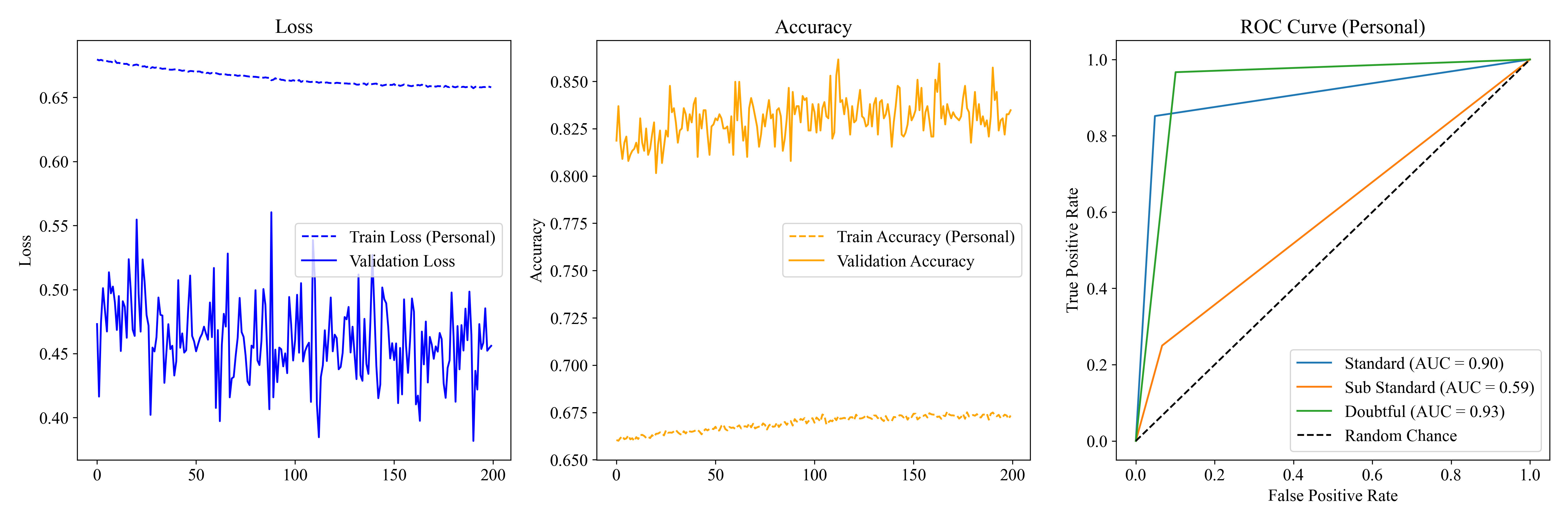} 
	\caption{Training and Validation Loss for Personal Loans}
	\label{fig:personal_loan_loss}
	\vspace{1em} 
	\begin{minipage}{0.7\textwidth} 
		This figure illustrates the training and validation loss curves for the personal loan model over the course of training.
		\begin{itemize}
			\item \textbf{Training Loss:} The curve representing the loss calculated on the training data set during each epoch of the model's training process. This helps in understanding how well the model is fitting to the training data.
			\item \textbf{Validation Loss:} The curve representing the loss calculated on the validation data set, which is used to assess how well the model generalizes to unseen data. A lower validation loss suggests better generalization.
			\item \textbf{Purpose of Comparison:} Monitoring both training and validation losses allows us to detect over fitting (if the training loss decreases while the validation loss increases) or under fitting (if both losses remain high).
		\end{itemize}
	\end{minipage}
\end{figure}
\vspace{1cm}
\noindent The training and validation loss curves for personal loans in fig\ref{fig:personal_loan_loss}  indicate that the training loss fluctuates between 0.657 and 0.659, indicating that the model is learning. Towards the end of the epochs, it shows a slight downward trend but stabilizes, suggesting a plateau. The validation loss ranges from 0.3818 (epoch 191) to 0.4977 (epoch 180), with a noticeable downward trend, particularly from epochs 191 to 194. The validation loss generally remains lower than the training loss, which is a positive indicator of the model's generalization ability. This trend suggests that the model can adapt well to unseen data. Overall, while the model is improving, there may still be opportunities for further optimization to enhance its performance. The observation of stabilization in training loss suggests that additional tuning may be required to achieve better results.\\

\begin{table}[h]
	\centering
	\begin{minipage}{0.8\textwidth}  
	\caption{Model Evaluation Metrics for Personal Loan Prediction. This table presents the performance of the predictive model across different classes (Standard, Sub Standard, and Doubtful), showing metrics like Precision, Recall, F1-Score, and Support. Additionally, the table includes aggregated performance metrics such as Accuracy, Macro Average, Weighted Average, and other key evaluation metrics like Train, Validation, and Test Accuracy, Precision, Recall, F1-Score, ROC AUC, and Cohen's Kappa. These metrics provide a comprehensive overview of model performance in predicting personal loan classes.}
	\begin{tabular}{|l|c|c|c|c|}
		\hline
		\textbf{Class} & \textbf{Precision} & \textbf{Recall} & \textbf{F1-Score} & \textbf{Support} \\ 
		\hline
		Standard       & 0.995968          & 0.851724        & 0.918216          & 870.00        \\ 
		Sub Standard   & 0.117647          & 0.250000        & 0.160000          & 32.00         \\ 
		Doubtful       & 0.241667          & 0.966667        & 0.386667          & 30.00         \\ 
		\hline
		\textbf{Accuracy}        & 0.834764 & 0.834764 & 0.834764 & 0.834764 \\ 
		\textbf{Macro Avg}      & 0.451760 & 0.689464 & 0.488294 & 932.00 \\ 
		\textbf{Weighted Avg}   & 0.941531 & 0.834764 & 0.875073 & 932.00 \\ 
		\hline
		\multicolumn{5}{|c|}{\textbf{Additional Metric}}  \\ 
		\hline
		\textbf{Train Accuracy}            & \multicolumn{4}{c|}{0.673375} \\ 
		\textbf{Validation Accuracy}       & \multicolumn{4}{c|}{0.834764} \\ 
		\textbf{Test Accuracy}             & \multicolumn{4}{c|}{0.834764} \\ 
		\textbf{Precision}                 & \multicolumn{4}{c|}{0.941531} \\ 
		\textbf{Recall}                    & \multicolumn{4}{c|}{0.834764} \\ 
		\textbf{F1-Score}                  & \multicolumn{4}{c|}{0.875073} \\ 
		\textbf{ROC AUC}                   & \multicolumn{4}{c|}{[0.9017, 0.5917, 0.932]} \\ 
		\textbf{Cohen’s Kappa}             & \multicolumn{4}{c|}{0.334187} \\ 
		\hline
	\end{tabular}
	\label{tab:model_metrics_personal}
\end{minipage}
\end{table}

\noindent The performance evaluation of the personal loan model in table\ref{tab:model_metrics_personal} reveals strong performance for the Standard Class, with a precision of 0.9959, indicating high confidence in predictions, although the recall of 0.8517 suggests some misclassification. The Sub Standard Class struggles significantly, achieving a precision of only 0.1176 and a recall of 0.2500, highlighting a major challenge in accurately identifying this class. The gls{f1}-score of 0.1600 for the Sub Standard Class further emphasizes its poor performance. In contrast, the Doubtful Class shows a high recall of 0.9667 but a low precision of 0.2417, leading to a moderate F1-score of 0.3867. Overall, the model's accuracy stands at 0.8348, which, while seemingly good, may be misleading due to class imbalances. The macro average precision of 0.4518 and recall of 0.6895 indicate suboptimal performance across classes, particularly due to the weak performance of the Sub Standard Class. However, the weighted average precision of 0.9415 suggests better performance when considering class sizes. Additional metrics show a train accuracy of 0.6734, indicating potential overfitting, while validation and test accuracies remain consistent at 0.8348. The \gls{roc} values indicate good discrimination for the Standard and Doubtful Classes, though the Sub Standard Class lags behind. Lastly, Cohen’s Kappa of 0.3342 indicates moderate agreement between predicted and actual classifications, signaling room for improvement in the model's predictive capabilities.

\subsection{Agriculture Loan Performance}

\begin{figure}[H]
	\centering
	\includegraphics[width=1.0\textwidth]{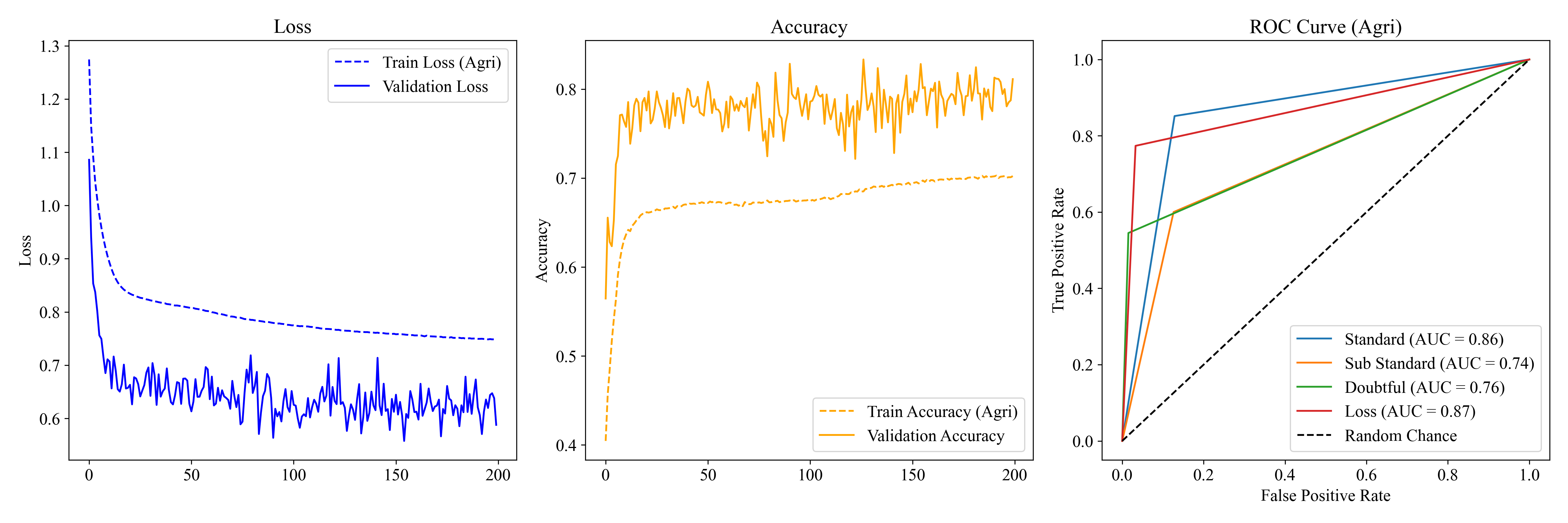} 
	\caption{Training and Validation Loss for Agriculture Loans}
	\label{fig:agriculture_loan_loss}
	\vspace{1em} 
	\begin{minipage}{0.8\textwidth}
		 This figure illustrates the training and validation loss curves for the agriculture loan model throughout the training process.
		\begin{itemize}
			\item \textbf{Training Loss:} This curve shows the loss calculated on the training dataset during each epoch of the model's training. It helps in tracking how well the model is learning from the training data.
			\item \textbf{Validation Loss:} This curve shows the loss calculated on the validation dataset, indicating how well the model is generalizing to unseen data. A lower validation loss suggests better model performance on new data.
			\item \textbf{Purpose of Comparison:} By comparing both the training and validation losses, we can detect overfitting (where the training loss decreases while the validation loss increases) or underfitting (where both losses remain high).
		\end{itemize}
	\end{minipage}
\end{figure}

\vspace{1cm}

\noindent The training and validation loss curves for agriculture loans in fig \ref{fig:agriculture_loan_loss} indicate that the training loss is consistently around 0.75, indicating some stagnation in learning. The accuracy is slightly improving, but the changes are minimal between epochs (moving from 0.6911 to 0.6954).  The validation loss fluctuates between epochs, indicating possible over fitting or instability in learning. Notably, some epochs show an increase in validation loss (from 0.5719 to 0.7138), which can be a concern.  There is an overall trend of improvement in accuracy, the fluctuations in validation metrics suggest the model may not be generalizing well to unseen data.  Some epochs show a notable drop in validation accuracy (from 0.7952 to 0.7281), which could signal over fitting, particularly if the training accuracy continues to improve. \\

\begin{table}[h]
	\centering
	\begin{minipage}{0.8\textwidth}
	\caption{Model Evaluation Metrics for Agricultural Loan Prediction. This table presents the performance of the predictive model across different classes (Standard, Sub Standard, and Doubtful), showing metrics like Precision, Recall, F1-Score, and Support. Additionally, the table includes aggregated performance metrics such as Accuracy, Macro Average, Weighted Average, and other key evaluation metrics like Train, Validation, and Test Accuracy, Precision, Recall, F1-Score, ROC AUC, and Cohen's Kappa. These metrics provide a comprehensive overview of model performance in predicting personal loan classes.}
	\begin{tabular}{|l|c|c|c|c|}
		\hline
		\textbf{Class} & \textbf{Precision} & \textbf{Recall} & \textbf{F1-Score} & \textbf{Support} \\ 
		\hline
		Standard       & 0.9753           & 0.8516        & 0.9092          & 3525.00      \\ 
		Sub Standard   & 0.0552           & 0.6000        & 0.1011          & 50.00        \\ 
		Doubtful       & 0.8286           & 0.5450        & 0.6576          & 488.00       \\ 
		Loss           & 0.2356           & 0.7735        & 0.3612          & 53.00        \\ 
		\hline
		\textbf{Accuracy}        & 0.8112 & 0.8112 & 0.8112 & 0.8112 \\ 
		\textbf{Macro Avg}      & 0.5237 & 0.6925 & 0.5073 & 4116.00 \\ 
		\textbf{Weighted Avg}   & 0.9372 & 0.8112 & 0.8625 & 4116.00 \\ 
		\hline
		\multicolumn{5}{|c|}{\textbf{Additional Metric}}  \\
		\hline
		\textbf{Train Accuracy}            & \multicolumn{4}{c|}{0.7022} \\ 
		\textbf{Validation Accuracy}       & \multicolumn{4}{c|}{0.8112} \\ 
		\textbf{Test Accuracy}             & \multicolumn{4}{c|}{0.8112} \\ 
		\textbf{Precision}                 & \multicolumn{4}{c|}{0.9372} \\ 
		\textbf{Recall}                    & \multicolumn{4}{c|}{0.8112} \\ 
		\textbf{F1-Score}                  & \multicolumn{4}{c|}{0.8625} \\ 
		\textbf{ROC AUC}                   & \multicolumn{4}{c|}{[0.8615, 0.7369, 0.764]} \\ 
		\textbf{Cohen’s Kappa}             & \multicolumn{4}{c|}{0.4578} \\ 
		\hline
	\end{tabular}
	\label{tab:model_metrics_agri}
\end{minipage}
\end{table} 

\noindent The Agricultural loan model achieves a reasonably high overall accuracy of 0.81, although this metric can be misleading due to class imbalance, with the majority of predictions dominated by the "Standard" class. Examining the macro average metrics (Precision 0.52, Recall 0.69, F1-Score 0.51) provides a more balanced view across classes, revealing that while the model has a decent recall (sensitivity to detecting classes), it struggles with precision, particularly for the minority "Sub Standard" class. The weighted average indicates better overall performance, heavily influenced by the "Standard" class’s high performance. \gls{roc} scores highlight further distinctions in class separability, with "Standard" achieving the highest AUC (0.86) and "Sub Standard" the lowest, signaling that additional tuning or feature engineering might enhance separation, especially for minority classes. Cohen's Kappa score of 0.46 suggests moderate agreement beyond random chance, indicating reasonable but improvable consistency. The model, though effective, has considerable scope for enhancement, particularly in refining predictions for "Sub Standard" and "Doubtful" classes \ref{tab:model_metrics_agri}.
\subsection{Economic Impact}
Implications of \gls{hyquc} in the Banking Sector:\\\\
The HyQuC-DeepNN-RTDPA model has been specifically tested for credit risk analysis, demonstrating its effectiveness in identifying complex patterns in loan applicant profiles and predicting creditworthiness. The model’s integration of quantum and classical deep learning layers allows it to manage diverse row types, such as various loan products (e.g., personal loans, mortgages, business loans), each with distinct risk profiles and data characteristics.\\

1. Credit Risk Analysis
The model is particularly suited for credit risk prediction due to its row-type-dependent structure. It can adapt to the unique characteristics of different customer groups, allowing for more granular and accurate risk assessments. Quantum layers, through operations like entanglement and quantum measurement, enhance the model’s ability to detect subtle correlations and interactions between features, such as income, credit history, and loan size. This leads to better risk discrimination, potentially reducing default rates and improving lending decisions.\\

2. Fraud Detection
Although initially applied to credit risk, the HyQuC-DeepNN-RTDPA model’s architecture is equally effective for other predictive tasks, such as fraud detection. By leveraging quantum-inspired entangling layers, the model can efficiently process high-dimensional transactional data to identify anomalous patterns indicative of fraudulent behavior. The ability to handle row-type-specific pre processing also allows the model to differentiate between transaction types (e.g., high-frequency vs. low-frequency) and apply custom detection techniques.\\

3. Predictive Maintenance and Forecasting
In addition to credit risk and fraud detection, the model is also well-suited for other predictive analysis purposes in the banking sector, such as: Customer churn prediction: By modeling different customer segments (e.g., high-value vs. low-value customers), the model can identify those at risk of leaving and suggest retention strategies.
Portfolio management: The quantum layers can capture the intricate relationships between financial assets, making the model useful for portfolio optimization and risk forecasting.
Operational risk management: By handling heterogeneous datasets, the model can help banks predict and mitigate operational risks related to internal processes and regulatory compliance.\\

4. Class Imbalance Handling
Another notable feature of the model is its ability to address class imbalance across different row types. This is particularly valuable in scenarios such as minority class loan applicants (e.g., applicants with non-traditional credit histories), where traditional models struggle. The HyQuC-DeepNN-RTDPA model effectively augments such datasets with quantum-inspired data augmentation techniques, improving model robustness and fairness. \\\\ Broader Applicability in Predictive Analysis
While this model has been applied to credit risk analysis, its hybrid architecture allows it to be extended to a wide range of predictive tasks across various domains. The ability to handle row-type-dependent processing makes it versatile for applications such as marketing analytics, customer behavior prediction, risk forecasting in other industries, and time-series analysis. Future work can explore these applications by adapting the row-type-specific pre processing and hybrid quantum-classical architecture to new data domains and problem contexts.

\section{Computational Constraints}

The experiments in this study were conducted under significant hardware limitations, which impacted certain aspects of the methodology and analysis.  The experiments were conducted on a system with 16 GB RAM, 8-core CPU specifically. The lack of access to high-performance computing infrastructure restricted our ability to perform computationally intensive operations such as the calculation of confidence intervals for key performance metrics (e.g., precision, recall, F1-score, accuracy, and ROC AUC). While these metrics are vital for providing a more robust and statistically sound evaluation, the hardware constraints necessitated a trade-off in prioritizing other critical analyses. \\\\ These limitations also influenced the scale of hyper-parameter tuning and the level of complexity in feature engineering and data augmentation that could be explored. As a result, the study relied on streamlined approaches and approximations to balance computational feasibility with the need for meaningful results. \\\\To address these constraints in future work, we propose the adoption of cloud-based computing solutions or access to dedicated high-performance computing facilities. Such resources will enable more comprehensive analyses, including the use of advanced statistical methods to calculate confidence intervals and other error estimates, thereby enhancing the robustness and reliability of the findings.

\section{Limitations of the Study}

\begin{enumerate}
	\item[] \textbf{Dimensionality Reduction Constraints:} The use of Principal Component Analysis  (\gls{pca}) for dimensionality reduction reduced the original feature set to only five components. While PCA aims to retain the most significant variance, this reduction may have resulted in the loss of critical information. As a result, the model’s ability to accurately differentiate between various loan statuses may have been compromised, particularly when subtle patterns essential for distinguishing between certain categories are lost. The limitations of current quantum simulators further restricted the depth of experimentation. Specifically, due to constraints in handling higher-dimensional data, which are currently unable to process datasets with up to 38 or 43 principal components, we were limited to testing our algorithm using only the first 5 principal components.  
	
	\item[] \textbf{Quantum Model Sensitivity:} Quantum models are inherently sensitive to the structure of the input data and the methods employed in data preprocessing. In this study, the reliance on PCA as the sole dimensionality reduction technique may not have fully captured the complexities of the dataset. Other dimensionality reduction methods or a more diversified preprocessing strategy could potentially improve the model’s performance by preserving more nuanced features of the data. Additionally, the results presented in this study were based on angle embedding. However, other embedding techniques, such as Amplitude Embedding,superposition encoding could be explored in future research. These techniques offer different ways of encoding classical data into quantum states and may provide a more effective representation for certain types of datasets. Furthermore, due to computational constraints, we were unable to conduct extensive parameter variations or perform a detailed sensitivity analysis, both of which could provide deeper insights into the model's behavior and help optimize its performance.
	
	\item[] \textbf{Imbalanced Class Distribution:} The dataset used in this study exhibited significant class imbalance, particularly in the 'Doubtful' and 'Loss' categories for both personal and agricultural loans. This imbalance can introduce bias in the model’s predictions, resulting in a performance metric that might favor the majority class. While accuracy may appear high, the model could under perform on minority classes, affecting the overall robustness and fairness of predictions. To address class imbalance,SMOTE (Synthetic Minority Over-sampling Technique) and its variants can be explored. These include \textbf{Borderline-SMOTE}, which focuses on generating synthetic samples near the decision boundary, and its further variants, \textbf{Borderline-SMOTE1} and \textbf{Borderline-SMOTE2}, which specifically target borderline or misclassified samples. \textbf{SMOTE-ENN} and \textbf{SMOTE-Tomek Links} combine SMOTE with data cleaning techniques like Edited Nearest Neighbor (ENN) and Tomek Links to refine the dataset after oversampling. \textbf{ADASYN (Adaptive Synthetic Sampling)} adjusts the number of synthetic samples based on the difficulty of minority class instances. \textbf{KMeans-SMOTE} uses clustering techniques to group similar minority class instances before applying SMOTE, ensuring that synthetic data respects the data structure. \textbf{SVMSMOTE} leverages Support Vector Machines (SVM) to create synthetic samples near the decision boundary, focusing on support vectors. Other variants, such as \textbf{Random-SMOTE}, generate synthetic samples by randomly interpolating between neighbors, while \textbf{Cluster-SMOTE} uses clustering algorithms to group similar data points before oversampling. These variants offer various strategies to improve the quality and relevance of the synthetic samples generated, depending on the specific characteristics of the dataset and problem at hand. However, due to computational constraints and the specific focus of this study, these techniques were not incorporated into our analysis. Future studies could investigate the impact of these balancing techniques on model performance, which may further enhance the model’s ability to handle imbalanced datasets and improve predictions for the minority classes.
	
	\item[] \textbf{Limited Hyperparameter Tuning:} Due to computational constraints, the hyperparameters of the quantum model may not have been thoroughly optimized. Hyperparameter tuning is a crucial step in improving the performance of machine learning models, and inadequate tuning may have limited the model’s potential to achieve its optimal performance. While a more extensive hyperparameter search could have potentially enhanced the model’s ability to capture underlying patterns more effectively, we did perform a structured hyperparameter tuning process as outlined previously.
	
	Beyond the hyperparameters we tuned, additional adjustments could be made in future work to further optimize the model’s performance. For instance:
	
	\begin{itemize}
		\item \textbf{Quantum Circuit Design:} Experimenting with different quantum circuit architectures, such as varying the number of layers, gates, and types of quantum gates, could influence model performance.
		\item \textbf{Learning Rate Schedules:} Implementing learning rate schedules (e.g., exponential decay, step decay) could improve convergence during training.
		\item \textbf{Optimizer Choices:} Testing different optimization algorithms (e.g., Adam, SGD, Adagrad) could result in more efficient training and potentially better performance.
		\item \textbf{Dropout Rates:} Implementing dropout or regularization techniques could help improve generalization by preventing overfitting.
		\item \textbf{Batch Normalization:} Incorporating batch normalization could help with stabilizing training by normalizing activations between layers.
		\item \textbf{Early Stopping:} Implementing early stopping criteria could prevent overfitting and save computational resources by halting training once performance plateaus.
	\end{itemize}
	
	Exploring these additional hyperparameters, along with more comprehensive searches, could enhance the robustness and effectiveness of the quantum model.

	\item[] \textbf{Interpretability of Results:} One of the significant challenges with quantum models is that their results are often difficult to interpret when compared to classical machine learning approaches. This lack of interpretability can pose a major limitation in practical applications, especially in sectors like banking, where understanding the reasoning behind model decisions is essential for ensuring trust and transparency. To address this, further research is needed to enhance the interpretability of quantum models, making them more suitable for real-world decision-making processes. Developing techniques that attribute quantum model predictions to specific features or components could improve transparency. For instance, methods such as SHAP (Shapley Additive Explanations) or LIME (Local Interpretable Model-Agnostic Explanations) could be adapted for quantum models, helping to explain the influence of various quantum features on the final outcome.

	\item[] \textbf{Generalizability:} The results of this study are based on a specific dataset and loan types, which may limit the generalizability of the findings to other datasets or loan categories. Variations in feature distributions, class representations, and domain-specific characteristics may lead to different outcomes if the model were applied to other contexts. Additional studies on diverse datasets would be needed to assess the broader applicability of the proposed approach.
	
	\item[] \textbf{Computational Limitations:} Current quantum computing hardware is constrained by several key limitations that affect its ability to scale effectively. First, the \textbf{number of qubits} available on quantum processors remains limited. While some systems feature a few dozen qubits, others may have up to a few hundred qubits. However, the full potential of these systems is constrained by issues such as limited qubit connectivity and the need for error correction, preventing them from being fully utilized for complex tasks.
	
	Second, \textbf{quantum coherence time}—the duration during which quantum states can remain in their superposition—remains short. This is problematic for running long or computationally intensive algorithms, as the system may lose information before completing the computation. The need for longer coherence times is critical for performing complex quantum machine learning tasks efficiently.
	
	Another limitation is \textbf{quantum error rates}. Current quantum computers are susceptible to noise and errors, which can significantly degrade the quality of computations. Quantum error correction techniques are still in development, and without these, the performance of quantum models on noisy devices may be unreliable, leading to inaccurate or suboptimal results.
	
	Furthermore, the \textbf{quantum-to-classical interface} (the interaction between quantum and classical components in hybrid models) remains a challenging aspect of scaling. The process of extracting useful results from quantum systems and combining them with classical computing resources can introduce inefficiencies, limiting the overall performance and scalability of quantum-enhanced models.
	
	Finally, \textbf{limited connectivity} between qubits on current quantum processors restricts the ability to perform arbitrary quantum operations efficiently. The topology of qubits and the need to establish quantum entanglements between distant qubits to perform certain operations add additional layers of complexity to model training.
	
	As quantum hardware continues to improve with advances in error correction, qubit scaling, and quantum coherence times, it is expected that these hardware limitations will be overcome, allowing for the development of more powerful quantum models capable of handling larger and more complex datasets in real-world applications.

	\item[] \textbf{Assumptions of \gls{pca}:} Principal Component Analysis (PCA) assumes that the principal components are orthogonal and that the dataset exhibits linear correlations among its features. These assumptions may not hold true for all datasets, particularly when complex non-linear relationships exist between features. If the data set does not conform to these assumptions, the dimensionality reduction process may lead to sub-optimal feature representations, potentially affecting model performance. Alternative techniques that account for non-linearity may be considered in future studies. Alternative techniques that account for non-linearity, such as t-Distributed Stochastic Neighbor Embedding (t-SNE), Kernel Principal Component Analysis (KPCA), Isomap, and Autoencoders, may be considered in future studies. These methods are better suited for capturing complex, non-linear relationships within the data, which could lead to improved feature representation and enhanced model performance.
\end{enumerate}

\section{Conclusion}

This framework presents a systematic approach for integrating QDL techniques into credit risk analysis, specifically designed for row-type dependent predictive analysis. It addresses essential theoretical foundations, practical implementation strategies, and explores the potential to enhance predictive capabilities within the banking sector. However, certain challenges remain, including the high computational costs associated with quantum processing, scalability limitations, and the current need for specialized hardware. Addressing these shortcomings will be essential for broader adoption and practical application in industry settings.

\section*{Appendix}
\addcontentsline{toc}{section}{Appendix}

\section{Model Functionality}To provide a clearer understanding of the interactions between the quantum and classical layers, a simple example is explored.

\subsection*{Data Points}
\begin{tabular}{|c|c|}
	\hline
	\textbf{Feature (X) } & \textbf{Label (Y)} \\
	\hline
	20 & True (1) \\
	\hline
	10 & False (0) \\
	\hline
\end{tabular}

\subsection{Step 1: Quantum Encoding}

\subsubsection*{Normalization:}
\begin{align*}
	\text{Normalized } x_1 & = \frac{20}{\sqrt{20^2 + 10^2}} = \frac{20}{\sqrt{500}} \approx 0.8944 \text{      and      }
	\text x_2 & = \frac{10}{\sqrt{20^2 + 10^2}} = \frac{10}{\sqrt{500}} \approx 0.4472
\end{align*}

The quantum states for each data point are represented as:

\begin{itemize}
	\item[] For Data Point 1 (20 True): $ |\psi_1\rangle = \begin{bmatrix} 0.8944 \\ 0 \end{bmatrix} $
	
	\item[] For Data Point 2 (10 False): $ |\psi_2\rangle = \begin{bmatrix} 0 \\ 0.4472 \end{bmatrix} $
\end{itemize}
\subsubsection{Quantum Processing}

Assuming we have a simple quantum circuit that applies a rotation gate $
R(\theta) = \begin{bmatrix}
	\cos\left(\frac{\theta}{2}\right) & -\sin\left(\frac{\theta}{2}\right) \\
	\sin\left(\frac{\theta}{2}\right) & \cos\left(\frac{\theta}{2}\right)
\end{bmatrix} $  to the state,  with $\theta$ = 0.325 radians.\\ 

$ \frac{\theta}{2} $ = 0.1625, $ \cos(0.1625) \approx 0.9877 $ and $\sin(0.1625) \approx 0.1616 $\\

The  rotation matrix: $ R_Z(0.325) \approx \begin{bmatrix} 	0.9877 & -0.1616 \\ 	0.1616 & 0.9877 \end{bmatrix} $ \\

Apply the Rotation to Each Quantum State\\

Quantum State for Data Point 1 (20, True):

Applying $R_Z(0.325)$: $
|\psi_{1,\theta}\rangle = R_Z(0.325)|\psi_1\rangle = \begin{bmatrix}
	0.9877 & -0.1616 \\ 0.1616 & 0.9877
\end{bmatrix} \begin{bmatrix} 0.8944 \\ 0 \end{bmatrix} = \begin{bmatrix}
	0.8845 \\ 0.1446 \end{bmatrix}$

Quantum State for Data Point 2 (10, False): 

Applying $R_Z(0.325)$: $|\psi_{2,\theta}\rangle = R_Z(0.325)|\psi_2\rangle = \begin{bmatrix}
	0.9877 & -0.1616 \\	0.1616 & 0.9877 \end{bmatrix} \begin{bmatrix} 0 \\ 0.4472 \end{bmatrix}  = \begin{bmatrix}
	-0.0723 \\	0.4417 \end{bmatrix} $

\subsection{Step 2: Classical Processing}

\subsubsection*{Feeding into Classical Layers:}

Assuming we have a simple dense layer with: Weights: $ w = [0.5, 0.5]$ (for two inputs), \\ $  Bias:  b = 0.1 $  and  Sigmoid:
$ \sigma(x) = \frac{1}{1 + e^{-x}}$\\

\noindent \textbf{Feature Selection} : In many machine learning models, specific features are selected to represent the data. The quantum state is represented as a vector with multiple components, such as $ | \psi \rangle = \begin{pmatrix} 0.8845 \\ 0.1446 \end{pmatrix}. $  Here, the dominant amplitude (in this case, \( 0.8845 \)) is often chosen as it may represent the most significant aspect of the quantum state. This is because the first amplitude can be linked to the probability of measuring a certain outcome, thus influencing the subsequent classical processing.\\

\noindent Output for Data Point 1: \\ $ \text{Output} = \sigma(w * 0.8845 + b) = \sigma(0.5 * 0.8845 + 0.1)  = \sigma(0.54225) = \frac{1}{1 + e^{-0.54225}} \approx 0.6321 $\\

\noindent Output for Data Point 2: $ \text{Output} = \sigma(w * 0.4417 + b) = \sigma(0.5 * 0.4417 + 0.1)  = \sigma(0.32215) \approx 0.5796 $

\subsection{Step 3: Loss Calculation}

Using Binary Cross-Entropy: $ \text{Loss} = - \left( y \cdot \log(\hat{y}) + (1 - y) \cdot \log(1 - \hat{y}) \right) $ \\

\noindent \text{For the first data point:} Assuming \(y = 1\) (True): \\ $ \text{Loss} = - \left( 1 \cdot \log(0.6321) + 0 \cdot \log(1 - 0.6321) \right) \approx -\log(0.6321) \approx 0.4587 $ \\
\noindent \text{For the second data point:} Assuming \(y = 0\) (False): \\$ \text{Loss} = - \left( 0 \cdot \log(0.5796) + 1 \cdot \log(1 - 0.5796) \right) \approx -\log(0.4204) \approx 0.8665 $\\

\noindent $ \text{Total Loss} = \frac{0.4587 + 0.8665}{2} \approx \frac{1.3252}{2} \approx 0.6626 $

\subsection{Backward Pass (Gradient Calculation)}

\subsubsection{Gradients for Classical Layers:}
To find the gradients with respect to the weights \( w \):

\noindent \textbf{For the first data point:} ${y=1}$
\[ \frac{\partial \text{Loss}}{\partial \hat{y}} = -\frac{y}{\hat{y}} + \frac{1 - y}{1 - \hat{y}} = -\frac{1}{0.6321} + 0 \approx -1.581 \]

\[ \frac{\partial \hat{y}}{\partial w} = \hat{y}(1 - \hat{y}) * x = 0.6321 * (1 - 0.6321) * 0.8845 \approx  0.2053 \]

\[ \frac{\partial \text{Loss}}{\partial w} = \frac{\partial \text{Loss}}{\partial \hat{y}} \cdot \frac{\partial \hat{y}}{\partial w} \approx -1.581 \cdot 0.2053 \approx -0.324 \]

\noindent \textbf{For the second data point:} ${y=0}$
\[ \frac{\partial \text{Loss}}{\partial \hat{y}} = -\frac{y}{\hat{y}} + \frac{1 - y}{1 - \hat{y}} = -\frac{0}{0.5796} + \frac{1}{1 - 0.5796} = 0 + \frac{1}{0.4204}\approx 2.376 \]

\[ \frac{\partial \hat{y}}{\partial w} = \hat{y}(1 - \hat{y}) \cdot x = 0.5796 \cdot (1 - 0.5796) \cdot 0.4417 \approx 0.1074 \]

\[\frac{\partial \text{Loss}}{\partial w} = \frac{\partial \text{Loss}}{\partial \hat{y}} \cdot \frac{\partial \hat{y}}{\partial w} \approx 2.376 \cdot 0.1074 \approx 0.255\]

\subsubsection{Gradients for Quantum Layers:}
To calculate the gradient of the output probabilities with respect to \( \theta \), we apply the parameter shift rule: $ \frac{\partial f(\theta)}{\partial \theta} = \frac{f\left(\theta + \frac{\pi}{2}\right) - f\left(\theta - \frac{\pi}{2}\right)}{2}. $  Here, \( f(\theta) \) is the probability of measuring the state corresponding to \( |\phi\rangle \).\cite{mitarai2018quantum} \cite{schuld2019evaluating}\\

\noindent Evaluate \( f\left(\theta + \frac{\pi}{2}\right) \):  First, we calculate the new output probabilities when \( \theta \) is increased by \( \frac{\pi}{2} \):
First, we calculate the new output probabilities when \( \theta \) is increased by \( \frac{\pi}{2} \):\\

\textbf{Step 1: Define the Parameter Shift Rotation Matrices}
For a rotation by \( \theta + \frac{\pi}{2} \):\[R\left(\theta + \frac{\pi}{2}\right) = \begin{bmatrix} 0 & -1 \\ 1 & 0 \end{bmatrix}\]
For \( f\left(\theta + \frac{\pi}{2}\right) \): \[ R\left(\theta - \frac{\pi}{2}\right) = \begin{bmatrix} 0 & 1 \\ -1 & 0 \end{bmatrix} \]

\textbf{Step 2: Apply \( R(\theta + \frac{\pi}{2}) \) and \( R(\theta - \frac{\pi}{2}) \) to Each Quantum State}

\paragraph{For Data Point 1}
1. Original quantum state:
\[| \psi \rangle = \begin{bmatrix} 0.8845 \\ 0.1446 \end{bmatrix}\]

2. Applying \( R(\theta + \frac{\pi}{2}) \):
\[| \psi_{\theta + \frac{\pi}{2}} \rangle = \begin{bmatrix} 0 & -1 \\ 1 & 0 \end{bmatrix} \cdot \begin{bmatrix} 0.8845 \\ 0.1446 \end{bmatrix} = \begin{bmatrix} -0.1446 \\ 0.8845 \end{bmatrix}\]

3. Applying \( R(\theta - \frac{\pi}{2}) \):
\[| \psi_{\theta - \frac{\pi}{2}} \rangle = \begin{bmatrix} 0 & 1 \\ -1 & 0 \end{bmatrix} \cdot \begin{bmatrix} 0.8845 \\ 0.1446 \end{bmatrix} = \begin{bmatrix} 0.1446 \\ -0.8845 \end{bmatrix}\]

\paragraph{For Data Point 2}
1. Original quantum state:
\[| \psi \rangle = \begin{bmatrix} -0.0723 \\ 0.4417 \end{bmatrix}\]

2. Applying \( R(\theta + \frac{\pi}{2}) \):
\[| \psi_{\theta + \frac{\pi}{2}} \rangle = \begin{bmatrix} 0 & -1 \\ 1 & 0 \end{bmatrix} \cdot \begin{bmatrix} -0.0723 \\ 0.4417 \end{bmatrix} = \begin{bmatrix} -0.4417 \\ -0.0723 \end{bmatrix}\]

3. Applying \( R(\theta - \frac{\pi}{2}) \):
\[| \psi_{\theta - \frac{\pi}{2}} \rangle = \begin{bmatrix} 0 & 1 \\ -1 & 0 \end{bmatrix} \cdot \begin{bmatrix} -0.0723 \\ 0.4417 \end{bmatrix} = \begin{bmatrix} 0.4417 \\ 0.0723 \end{bmatrix}\]

\textbf{Step 3: Calculate Probabilities for Each Adjusted State} \\

\noindent To find the probabilities for each adjusted state \( | \psi_{\theta + \frac{\pi}{2}} \rangle \) and \( | \psi_{\theta - \frac{\pi}{2}} \rangle \), square the amplitudes (assuming measurement in the computational basis).

\textbf{For Data Point 1}
\[f\left(\theta + \frac{\pi}{2}\right) = (-0.1446)^2 = 0.0209\]
\[f\left(\theta - \frac{\pi}{2}\right) = (0.1446)^2 = 0.0209\]

\textbf{For Data Point 2}
\[f\left(\theta + \frac{\pi}{2}\right) = (-0.4417)^2 = 0.1951\]
\[f\left(\theta - \frac{\pi}{2}\right) = (0.4417)^2 = 0.1951\]

\textbf{Step 4: Use Parameter Shift Rule for Gradient Calculation}

Applying the parameter shift rule for each data point:

\[\frac{\partial f(\theta)}{\partial \theta} = \frac{f\left(\theta + \frac{\pi}{2}\right) - f\left(\theta - \frac{\pi}{2}\right)}{2}\]

For both data points, since the probabilities are symmetric, the gradient is zero:

\[\frac{\partial f(\theta)}{\partial \theta} \approx 0\]\\

\noindent This result indicates no change in probability with a small shift in \( \theta \), implying that the gradient is zero at this point.\\

\noindent \textbf{Step 6: Parameter Updates}\\

\noindent Assuming a learning rate \( \eta = 0.01 \): \\

\noindent \textbf{Update Weights for Classical Layer}\\

\noindent Update Weights for Classical Layer:  $ w(t + 1) = w(t) - \eta \cdot \frac{\partial \text{Loss}}{\partial w} $
with initial \( w = 0.5 \).\\

\noindent \textbf{Data Point 1}
Given: $\frac{\partial \text{Loss}}{\partial w} \approx -0.324 $, Updating \( w \) for Data Point 1: $w(1) = 0.5 - 0.01 \cdot (-0.324) = 0.5 + 0.00324 = 0.50324 $ \\

\noindent \textbf{Data Point 2}
Given: $ \frac{\partial \text{Loss}}{\partial w} \approx 0.255 $, Updating \( w \) for Data Point 2: $ w(2) = 0.5 - 0.01 \cdot 0.255 = 0.5 - 0.00255 = 0.49745 $ \\

\textbf{Summary of Updates}
After processing each data point, the updated weights are:
\begin{itemize}
	\item[] After Data Point 1: \( w(1) = 0.50324 \)
	\item[] After Data Point 2: \( w(2) = 0.49745 \)
\end{itemize}

\noindent At the end of Epoch 1, the model has adjusted both classical and quantum parameters based on the computed gradients, improving its performance for the next round of training.

\printglossary[type=\acronymtype, title=List of Abbreviations]

\section*{Declarations}

\subsection*{Conflict of Interest Statement}
The authors declare that there is no conflict of interest regarding the publication of this manuscript.

\subsection*{Ethical Approval and Consent to Participate}
Not applicable. This study did not involve any human or animal subjects requiring ethical approval.

\subsection*{Consent for Publication}
Not applicable. This manuscript does not contain any individual person’s data in any form.

\section*{Data Availability Statement}
The datasets contains personal information of borrowers from national bank, hence cannot be shared.

\subsection*{Competing Interests/Authors' Contributions}
The authors declare that they have no competing interests.

\subsection*{Authors' Contributions:}
All authors contributed equally.

\subsection*{Funding}
This research received no specific grant from any funding agency in the public, commercial, or not-for-profit sectors.

\bibliographystyle{unsrt}
\bibliography{BankQuantum}

\begin{longtable}{|c|l|p{8cm}|}
			\caption{Column Description} \label{tab:column_description} \\
			\hline
			\textbf{COLUMN NO} & \textbf{COLUMN NAME} & \textbf{DESCRIPTION} \\
			\hline
			\endfirsthead
			
			\multicolumn{3}{c}%
			{{\tablename\ \thetable{} -- continued from previous page..}} \\
		
			\hline	
			\textbf{COLUMN NO} & \textbf{COLUMN NAME} & \textbf{DESCRIPTION} \\
			\hline
			\endhead
			
			\hline \multicolumn{3}{r}{{Continued on next page}} \\ \hline
			\endfoot
			
			\hline
			\endlastfoot
		
		1 & Q & Quarter \\
		2 & BRCD & Branch Code \\
		3 & CUSTID & Customer ID \\
		4 & ACCTID & Account ID \\
		5 & SEGCD & Segment Code \\
		6 & ORGCD & Org. Code \\
		7 & STFCD & Staff Code \\
		8 & RESFLG & Resident Flag \\
		9 & BKGSINCEDT & Date since when banking with the Bank \\
		10 & GENCD & Gender Code \\
		11 & BIRTHDT & Date of Birth \\
		12 & MARST & Marital Status \\
		13 & OCUCD & Occupation Code \\
		14 & DRYLAND & Area of DRY Land \\
		15 & WETLAND & Area of WET Land \\
		16 & SBWCLMTAMT & SBI Working Capital Limit \\
		17 & PARTBKFLG & Participating Bank Flag \\
		18 & SBTLLMTAMT & SBI Term Loan Limit \\
		19 & OPINIONDT & Opinion Date \\
		20 & TPGAMT & Third Party Guarantee Amount \\
		21 & BORWORAMT & Net Worth of the Borrower \\
		22 & PRODCODE & Product Code \\
		23 & FACCD & Facility Code \\
		24 & SUBFACCD & Sub Facility Code \\
		25 & PRIFLG & Priority Flag \\
		26 & DIRFINFLG & Direct Finance Flag \\
		27 & SECTORCD & Sector Code \\
		28 & SCHEMECD & Scheme Code \\
		29 & ACTCD & Account ID \\
		30 & SANCTIONDT & Date of Sanction \\
		31 & SANAUTCD & Sanctioning Authority \\
		32 & OPENINGDT & Date of Opening \\
		33 & LIMITAMT & Limit Amount \\
		34 & DOCREVDT & Document Revival Date \\
		35 & FOODNONFLG & Food or Non-Food Sector Flag \\
		36 & INTRATE & Rate of Interest \\
		37 & OUTAMT & Amount Outstanding \\
		38 & PRISECCD1 & Primary Security Code 1 \\
		39 & PRISECCD2 & Primary Security Code 2 \\
		40 & PRISECAMT & Primary Security Amount \\
		41 & SPLCSAMT & Specific Collateral Security Amount \\
		42 & INCAMT & Interest Not Collected Amount \\
		43 & DISTTCD & District Code \\
		44 & POPCD & Population Code \\
		45 & UNIFUNFLG & Unit Function Status Flag \\
		46 & ACCSTACD & Accounting Std Code - Previous Quarter \\
		47 & ORIGINAMT & Original Amount \\
		48 & RENEWALDT & Date of Renewal \\
		49 & ALLCUSTID & Customer ID \\
		50 & DPAMT & Drawing Power Amount \\
		51 & INSEXPDT & Date of Insurance Expiry \\
		52 & INSAMT & Insurance Amount (Sum Assured) \\
		53 & FULDISFLG & Loan Fully Disbursed? \\
		54 & LASCREDT & Date of Last Credit \\
		55 & REPTYPCD & Repayment Type Code \\
		56 & PERIOD & Frequency of Interest Application \\
		57 & YTDINTAMT & Year-to-date interest applied \\
		58 & YTDCSUMAMT & Year-to-date credit summation \\
		59 & QDISMADAMT & YTD Disbursement (i.e. During the Year) \\
		60 & NUMOFINST & No of Instalments \\
		61 & REPFRQCD & Repayment Frequency Code \\
		62 & FIRINSDT & Date of First Instalment \\
		63 & INSTALAMT & Instalment Amount \\
		64 & RECAMT & INCA Recovered up to the Quarter \\
		65 & CALC\_ASSET & Calculated Asset (CALC) \\
		66 & TFRDT & Date of Transfer to Recalled Assets \\
		67 & REASONCD & Reason for Transfer to Recalled Assets \\
		68 & GINAMT & Amount originally Transferred to RA \\
		69 & RECALLDT & Date of Recall \\
		70 & SUTFILAMT & Suit Filed Amount \\
		71 & CUSTTYPE & Customer Type \\
		72 & RETAINAMT & DICGC/ECGC/CGTSI claim Retainable Amount \\
		73 & WOSACD & Retaining Amount \\
		74 & CUSTOTLMT & Customer Total Limit (CALC) \\
		75 & CUSTOTOUT & Customer Total Outstanding (CALC) \\
		76 & SUBSIDYAMT & Subsidy received and held amount \\
		77 & NFMRGAMT & Out of Total Security, Cash Security \\
		78 & MAR\_CALC & Calculated Asset as on March (Last) \\
		79 & MAR\_URIPY & URIPY as on March (Last) \\
		80 & PROV\_TOTAL & Current Provision Amount \\
		81 & IRAC & 1 - Std, 2 Sub-Std, 3-Doubtful, 4 Loss \\
	\end{longtable}

\section*{Author Information}
Authors and Affiliations:\\
\textbf{PhD, Analytics and  Decision Science}, IIM Mumbai, India. {minati.rath.2019@iimmumbai.ac.in}.\\
\textbf{Professor, Analytics and Decision Science}, IIM Mumbai, India.
{hemadate@iimmumbai.ac.in}.
\section*{Corresponding Authors}
Correspondence to {minati06@gmail.com}{, Minati Rath} 
	
\end{document}